\documentclass[preprint,aps,prd,showpacs,showkeys,nofootinbib,
preprintnumbers,superscriptaddress,tightenlines]{revtex4-1}
\usepackage{amsmath,amsfonts,amssymb,amscd,amsxtra,amsthm}
\usepackage{graphicx}  % Include figure files
\usepackage{epstopdf}
\usepackage{color}
\usepackage{dcolumn}  % Align table columns on decimal point
\usepackage{bm}          % bold math
\usepackage{multirow}
\usepackage{slashed}
\usepackage[utf8]{inputenc}
\usepackage[normalem]{ulem} % \sout{old text} for strikeout 
\usepackage[dvipsnames]{xcolor} % For blue in-text comments and
\usepackage{array}
\usepackage{slashed}
\renewcommand\sout{\bgroup \color{red} \ULdepth=-.5ex \ULset}

\newcommand\mprdel{\bgroup \color[rgb]{0.8,0.1,0.8} \ULdepth=-.5ex \ULset}
\begin{document}

\preprint{INHA-NTG-05/2017}
\title{Strong decays of exotic and nonexotic heavy baryons \\
in the chiral
quark-soliton model}

\author{Hyun-Chul Kim}
\affiliation{Department of Physics, Inha University, Incheon 22212,
  Republic of Korea}
\email{hchkim@inha.ac.kr}

\affiliation{School of Physics, Korea Institute for Advanced Study
  (KIAS), Seoul 02455, Republic of Korea}

\author{Maxim V. Polyakov}
\affiliation{Institut f\"ur Theoretische Physik II, Ruhr-Universit\"at
  Bochum, D--44780 Bochum, Germany} 
\email{maxim.polyakov@tp2.ruhr-uni-bochum.de}

\affiliation{Petersburg Nuclear Physics Institute, Gatchina,
  St. Petersburg 188 300, Russia}

\author{Micha{\l} Prasza{\l}owicz}
\affiliation{M. Smoluchowski Institute of Physics, Jagiellonian
  University, {\L}ojasiewicza 11, 30-348 Krak{\'o}w, Poland}
\email{michal.praszalowicz@uj.edu.pl}

\author{Ghil-Seok Yang}
\affiliation{Department of Physics, Soongsil University, Seoul 06978,
  Republic of Korea}\email{ghsyang@ssu.ac.kr}

\begin{abstract}
In the large $N_c$ limit both heavy and light baryons are described by
the universal mean field, {which} allows us 
to relate the properties of heavy baryons to light ones. 
With the only input from the decays of {\it light} octet baryons (due
to the universality of the chiral mean field),  
excellent description of strong decays of {\it both} charm and bottom
sextets is obtained.  The parameter-free prediction for the widths of exotic
antidecapentaplet ($\overline{\boldsymbol{15}}$) baryons is 
also made. The exotic heavy baryons should be anomalously  
narrow despite of large phase space available. In particular, the
widths of $\Omega_c(3050)$ and $\Omega_c(3119)$,  
interpreted as members of $\overline{\boldsymbol{15}}$-plet, 
are very small: 0.48~MeV and 1.12~MeV respectivly. This result is in
very good agreement with the measurements of the LHCb Collaboration
and provides natural and parameter-free explanation 
of the LHCb observation that $\Omega_c(3050)$ and $\Omega_c(3119)$
have anomalously small widths among five recently observed states. 
\end{abstract}



\maketitle

\section{Introduction}

In {a} recent paper \cite{Kim:2017jpx} we have proposed
to interpret two narrow $\Omega_c$ resonances  reported  recently by
the LHCb Collaboration \cite{Aaij:2017nav} as exotic pentaquark states
belonging to the SU(3) representation $\overline{\bf 15}$
(antidecapentaplet) \cite{Diakonov:2010tf}. To this end we have used the chiral quark
soliton model ($\chi$QSM) \cite{Diakonov:1987ty} (for  a review see
Ref.~\cite{Christov:1995vm} and references therein) modified in the
spirit of heavy quark symmetry~\cite{Isgur:1989vq, Isgur:1991wq,
  Georgi:1990um} to accommodate one heavy quark
\cite{Diakonov:2010tf,Yang:2016qdz}. {The} $\chi$QSM is based on
an old argument by Witten~\cite{Witten:1979kh}, which says that in the
limit of a large number of colors ($N_c \rightarrow \infty$), $N_{c}$
relativistic valence quarks generate chiral mean fields represented by
a distortion of a Dirac sea that in turn interact with the valence
quarks themselves.  In this way a self-consistent configuration called
a {soliton} is formed. The mean fields exhibit so called {\em
  hedgehog} symmetry, which means that 
neither 
{quark spin ($\bm{S}_q$) nor quark isospin ($\bm{T_q}$)  are "good" quantum
numbers. Instead a {\em grand spin} $\bm{K}=\bm{S}_q+\bm{T}_q$ is a good
quantum number.  

In order to project out spin and isospin quantum numbers one has to
rotate the  soliton, both in flavor and configuration
spaces. These rotations are then subsequently quantized
semiclassically and the collective Hamiltonian  is computed. 
The model predicts rotational baryon spectra that satisfy the
following selection rules: 

\begin{itemize}
\item allowed SU(3) representations must contain states with hypercharge
$Y^{\prime}=N_{c}/3$,
\vspace{-0.25cm}
\item the isospin $\bm{T}^{\prime}$ of the states with $Y^{\prime}%
=N_{c}/3$ couples with the soliton spin $\bm{J}$ to a singlet:
$\bm{T}^{\prime}+\bm{J}=0$.
\end{itemize}
In the case of light positive parity baryons the lowest allowed
representations are $\mathbf{8}$ of spin 1/2, $\mathbf{10}$ of spin
3/2, and also exotic pentaquark representation
$\overline{\mathbf{10}}$ of spin 1/2 with the lightest state
corresponding to the putative $\Theta^{+}(1540)$. 
Chiral models in general predict that pentaquarks are light 
\cite{Praszalowicz:2003ik,Diakonov:1997mm}
and -- in some specific models -- narrow \cite{Diakonov:1997mm}.
\begin{figure}[htp]
\centering
\includegraphics[width=9cm]{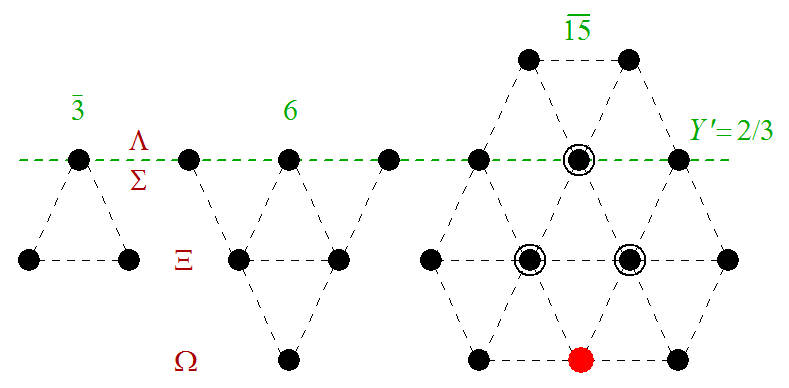} \vspace{-0.1cm}
\caption{Lowest lying SU(3) flavor representations allowed by the
  constraint $Y'=2/3$. The first  
{\em exotic} representation, $\overline{\mathbf{15}}$ contains the
putative pentaquark states $\Omega_c$ 
with $\Omega_c^0$ marked in red.}
\label{fig:irreps2}%
\end{figure}

In order to construct a heavy baryon in the $\chi$QSM we have to strip
off one light quark from the valence level and quantize the soliton
with a new constraint $Y^{\prime}=(N_{c}-1)/3$,
which modifies the above selection rules in the following way:
\begin{itemize}
\item allowed SU(3) representations must contain states with hypercharge
$Y^{\prime}=(N_{c}-1)/3$,
\vspace{-0.25cm}
\item the isospin $\bm{T}^{\prime}$ of the states with $Y^{\prime}%
=(N_{c}-1)/3$ couples with the soliton spin $\bm{J}$ to a singlet:
$\bm{T}^{\prime}+\bm{J}=0$.
\end{itemize}

This $N_c-1$ light
quark configuration is then coupled to a heavy quark $Q$ to form a
color singlet. The lowest allowed SU(3) representations are shown in
Fig.~\ref{fig:irreps2}. They correspond to the soliton in
representation in $\overline{\mathbf{3}}$ of spin 0  and to
${\mathbf{6}}$ of spin 1. Therefore the baryons constructed  from such
a soliton and a heavy quark form an SU(3) antitriplet of spin 1/2 and
two sextets  of spin 1/2 and 3/2 that are subject to a hyperfine 
splitting. The next allowed representation of the rotational excitations
corresponds to the exotic $\overline{\mathbf{15}}$ of spin 0 or spin
1 \cite{Kim:2017jpx}. The spin 1 soliton has  lower mass
and when it couples with a heavy quark it forms spin 1/2 or 3/2 exotic 
multiplets that should be hyperfine split similarly to the ground
state sextets by $\sim 70$~MeV. In Ref.~\cite{Kim:2017jpx} we have
proposed to interpret two LHCb states: $\Omega^0_c(3050)$ and
$\Omega^0_c(3119)$ as $1/2^+$ and $3/2^+$ pentaquarks  belonging to
the SU(3) $\overline{\mathbf{15}}$. As can be seen from
Fig.~\ref{fig:irreps2} they belong to the isospin triplets, and
therefore should have charged partners of the same mass, which allows,
in principle, for rather straightforward experimental verification of
this interpretation.

In the present paper we calculate strong decay widths of nonexotic and
exotic heavy quark baryons (both charm and bottom) in an approach
proposed many years ago by Adkins, Nappi and
Witten~\cite{Adkins:1983ya} and expanded in
Ref.~\cite{Diakonov:1997mm}, which is based on the Goldberger-Treiman 
relation where strong decay constants are expressed in terms of the
axial-vector current couplings. We  show that by fixing  
these axial-vector current couplings from the hyperon decays in
the light sector, we can predict strong decay widths of heavy baryons  
with only one free parameter related to the modification of these
couplings due to the fact that the soliton is constructed from $N_c-1$
rather than $N_c$ light quarks. We test our approach against
experimentally known charm and bottom  baryon decay 
widths, and then show that the decay widths of $\Omega^0_c(3050)$ and
$\Omega^0_c(3119)$ are small and compatible with the LHCb
measurements. Overall agreement of the predicted decay widths with
experiment is more than satisfactory and exceeds the expected model
accuracy, which is believed to be at the level of 10\% -- 20\%. 

The rotational states described above  correspond to positive
parity. Negative parity states generated by the soliton configurations
with one light  quark excited to the valence level from the Dirac sea
have been discussed in Ref.~\cite{Kim:2017jpx}. 
It has been shown that the remaining three LHCb $\Omega^0_c$ states
can be accommodated in such an approach. The formalism used in the
present paper has to be rather strongly modified to describe negative
parity state decay widths, and we plan to address this issue in a
separate publication. 

Other authors have also considered a possibility that at least some
of the LHCb $\Omega_c$ states may be interpreted as
pentaquarks~\cite{Yang:2017rpg, Huang:2017dwn, An:2017lwg,
  Anisovich:2017aqa} in different variants of the quark (or
quark-diquark) models. However, the states considered in these papers 
have negative parity and are isospin singlets. A simpler
interpretation that they are $p$-wave or 
radial and $p$-wave excitations of the $ss$-diquark $c$-quark system
has been put forward in Refs.~\cite{Karliner:2017kfm, Wang:2017hej,
  Wang:2017vnc,Cheng:2017ove,Chen:2017gnu,Ali:2017wsf}. 
Both radial and $p$-wave excitations have been also studied in the
framework of the QCD sum rules~\cite{Agaev:2017jyt,
  Chen:2017sci,Wang:2017zjw,Aliev:2017led,Azizi:2017bgs}, in a
phenomenological approach \cite{Zhao:2017fov} and on the lattice
\cite{Padmanath:2017lng}. A nonperturbative holographic QCD has been
applied to investigate heavy, regular, and exotic baryons in
Refs.~\cite{Liu:2017xzo}. The variety of quantum number assignments
proposed in the above references has been possible because  all the
above approaches (including the one in the present paper) suffer from
systematic uncertainties that exceed tiny experimental errors of the
heavy baryon masses. An ongoing analysis of spin and parity of the
LHCb $\Omega_c$ states is therefore of utmost importance to
discriminate different theoretical models. 

The paper is organized as follows. In Sec.~\ref{sec:genformalism}
we describe briefly the formalism for description of the baryon strong
decays in the $\chi$QSM. In the following
Sec.~\ref{sec:sextetdecays} we test the mean field picture of heavy
baryons against the known strong decays of ground-state baryon
sextets. In Sec.~\ref{sec:a15decays} the parameter-free prediction
for decays of exotic $\overline{\boldsymbol{15}}$-plet baryons is
made. Particular attention is paid to strong decays of
$\Omega_c(3050)$ and $\Omega_c(3119)$ which we interpret 
as  members of the exotic antidecapentaplet. Conclusions are presented
in Sec.~\ref{sec:con}.

\section{Heavy baryons and their strong decays in the chiral quark-soliton model }
\label{sec:genformalism}

In our picture the $\overline{\boldsymbol{3}}$, sextet and exotic
$\overline{\boldsymbol{15}}$ baryons shown in Fig.~\ref{fig:irreps2}
are rotational excitations of  the meson mean field, which is
essentially the same  as for light baryons.   
The corresponding 
wave function of the light sector is
given in terms of the Wigner rotational $D(A)$ matrices
\begin{align}
\Psi_{(B\,;\,-Y^{\prime}\,S\,S_{3})}^{(\mathcal{R\,}%
;\,B)}(A)  &  =\langle A|\mathcal{R},\,B,\,(-Y^{\prime},S,S_{3})\rangle
\nonumber\\
&  =\sqrt{\text{dim}(\mathcal{R})}\,(-)^{S_{3}-Y^{\prime}/2}D_{(Y,\,T,\,T_{3}%
)(Y^{\prime},\,S,\,-S_{3})}^{(\mathcal{R})\ast}(A) \label{eq:rotwf}%
\end{align}
where $\mathcal{R}$ denotes the SU(3) representation of the light sector,
$B=(Y,T,T_{3})$, stands for the SU(3) quantum numbers of a baryon in question,
and the second index of the $D$ function, $(Y^{\prime},S,-S_{3})$, corresponds
to the soliton spin. For the heavy baryons $Y^\prime=(N_c-1)/3$. $A(t)$ denotes
relative {\em configuration space} -- SU(3) {\em group space} rotation matrix.

The total wave function of a heavy baryon of spin $J$ is
constructed by coupling (\ref{eq:rotwf}) to a heavy quark ket $\left\vert
1/2,s_{3}\right\rangle $, with a pertinent SU(2) Clebsch-Gordan coefficient:
\begin{equation}
|\mathcal{R},\,B,\,J,J_{3}\rangle=%
%TCIMACRO{\dsum \limits_{S_{3},s_{3}}}%
%BeginExpansion
{\displaystyle\sum\limits_{S_{3},s_{3}}}
%EndExpansion
\left(
\begin{array}
[c]{cc}%
S & 1/2\\
S_{3} & s_{3}%
\end{array}
\right\vert \left.
\begin{array}
[c]{c}%
J\\
J_{3}%
\end{array}
\right)  \,\left\vert 1/2,s_{3}\right\rangle
\,|\mathcal{R},\,B,\,(-Y^{\prime 
},S,S_{3})\rangle\label{eq:fullwf}%
\end{equation}

The masses of various multiplets are obtained by sandwiching the
collective Hamiltonian between the wave function (\ref{eq:fullwf});
see for details Refs.~\cite{Kim:2017jpx,Yang:2016qdz}. To calculate
the decays of the heavy baryons one has to sandwich the corresponding  
decay operator between the wave functions (\ref{eq:fullwf}).
Following Ref.~\cite{Diakonov:1997mm} we  use in this paper the
decay operator describing the emission of a $p$-wave pseudoscalar
meson $\varphi$, as in the case of regular baryons, with possible
rescaling of the coefficients $G_{i}$ (see below),
\begin{equation}
\mathcal{O}_{\varphi}=\frac{3}{M_{1}+M_{2}}
\sum_{i=1,2,3} \left[  G_{0}D_{\varphi\,i}^{(8)}-G_{1}\,d_{ibc}%
D_{\varphi\,b}^{(8)}\hat{S}_{c}-G_{2}\frac{1}{\sqrt{3}}D_{\varphi\,8}%
^{(8)}\hat{S}_{i}\right]  p_{i} . \label{eq:dec-op}%
\end{equation}
We are considering decays $B_{1}\rightarrow B_{2}+\varphi$, 
where $M_{1,2}$ denote masses of the
initial and final baryons respectively and $p_{i}$ is the
c.m. momentum of the outgoing meson of mass 
$m$:%
\begin{equation}
\left\vert \vec{p}\, \right\vert =p=\frac{1}{2M_{1}}\sqrt{(M_{1}^{2}%
-(M_{2}+m)^{2})(M_{1}^{2}-(M_{2}-m)^{2})}%
\end{equation}

The decay width is  related to the matrix element of
$\mathcal{O}_{\varphi }$ squared, summed over the final and averaged
over the initial spin and isospin  denoted as $\overline{\left[ 
\ldots\right]  ^{2}}$;  see the appendix of
Ref.~\cite{Diakonov:1997mm} for details of the corresponding
calculations,
\begin{equation}
\Gamma_{B_{1}\rightarrow B_{2}+\varphi}=\frac{1}{2\pi}\overline{\left\langle
B_{2}\left\vert \mathcal{O}_{\varphi}\right\vert B_{1}\right\rangle ^{2}%
}\,\frac{M_{2}}{M_{1}}p.
\end{equation}
Here  factor $M_{2}/M_{1}$ is the same as in heavy baryon chiral
perturbation theory (HBChPT); see {\em e.g.}, Ref.~\cite{Cheng:2006dk}. 
This factor arises because in the heavy quark effective theory and in
HBChPT the velocities of $B_1$ and $B_2$ are the same up 
to corrections of order $O(1/m_Q)$, which we neglect.

The pseudoscalar meson-baryon couplings can be related to the
transition ($B_1\to B_2$) axial-vector constants with the help of
the Goldberger-Treiman relation, see Ref.~\cite{Yan:1992gz} for the 
derivation in the case of heavy baryons. Using the Goldberger-Treiman
relation we obtain for the couplings $G_{0,1,2}$ 
\begin{equation}
\left\{  G_{0},G_{1},G_{2}\right\}  =\frac{M_{1}+M_{2}}{2F_{\varphi}}\frac
{1}{3}\left\{  -a_{1},a_{2},a_{3}\right\},
\end{equation}
where constants\footnote{
For the reader's convenience we give the relations of the constants
$a_{1,2,3}$ to nucleon axial charges in the chiral limit:  
$g_{A}=\frac{7}{30}\left(
  -a_{1}+\frac{1}{2}a_{2}+\frac{1}{14}a_{3}\right)$, $g_A^{(0)}=\frac
12 a_3$,  
$g_{A}^{(8)}=\frac{1}{10 \sqrt 3}\left(
  -a_{1}+\frac{1}{2}a_{2}+\frac{1}{2}a_{3}\right)$} 
$a_{1,2,3}$ enter the definition of the axial-vector
current~\cite{Praszalowicz:1998jm} and have been extracted from the
semileptonic decays of  the
  baryon octet in Ref.~\cite{Yang:2015era},
\begin{equation}
a_{1}=-3.509\pm0.011,\;a_{2}=3.437\pm0.028,\;a_{3}=0.604\pm0.030.
\label{eq:a123}%
\end{equation}
For the decay constants $F_{\varphi}$ we have chosen the convention in which
$F_{\pi}=93$~MeV and $F_{K}=1.2\,F_{\pi}=112$ MeV.

The final formula for the decay width in terms of axial-vector
constants $a_{1,2,3}$ reads as follows:% 
\begin{align}
\label{eq:Gammageneral}
\Gamma_{B_{1}\rightarrow B_{2}+\varphi}  &  =\frac{1}{72\pi}\frac{p^{3}%
}{F_{\varphi}^{2}}\frac{M_{2}}{M_{1}}G_{\mathcal{R}_{1}\rightarrow
\mathcal{R}_{2}}^{2}\times\\
&  \times3\frac{\dim\mathcal{R}_{2}}{\dim\mathcal{R}_{1}}\left[
\begin{array}
[c]{cc}%
\boldsymbol{8} & \mathcal{R}_{2}\\
01 & Y^{\prime}S_{2}%
\end{array}
\right\vert \left.
\begin{array}
[c]{c}%
\mathcal{R}_{1}\\
Y^{\prime}S_{1}%
\end{array}
\right]  ^{2}\left[
\begin{array}
[c]{cc}%
\boldsymbol{8} & \mathcal{R}_{2}\\
Y_{\varphi}T_{\varphi} & Y_{2}T_{2}%
\end{array}
\right\vert \left.
\begin{array}
[c]{c}%
\mathcal{R}_{1}\\
Y_{1}T_{1}%
\end{array}
\right]  ^{2}\nonumber
\end{align}
Here $\mathcal{R}_{1,2}$ are the SU(3) representations of the initial and final baryons, $[.. .|..]$ 
are the SU(3) isoscalar factors.
The decay constants $G_{\mathcal{R}_{1}\rightarrow
\mathcal{R}_{2}}$ are calculated from the matrix elements of
(\ref{eq:dec-op})  as
\begin{align}
\overline{\boldsymbol{15}}_{1}  &  \rightarrow\overline{\boldsymbol{3}}%
_{0}\qquad G_{\overline{\boldsymbol{3}}}=-a_{1}-\frac{1}{2}a_{2}=0.44, \notag \\
\overline{\boldsymbol{15}}_{1}  &  \rightarrow\boldsymbol{6}_{1}\qquad
G_{\boldsymbol{6}}=-a_{1}-\frac{1}{2}a_{2}-a_{3}=-0.16,\notag \\
\boldsymbol{6}_{1}  &  \rightarrow\overline{\boldsymbol{3}}_{0}\qquad
H_{{\overline{\boldsymbol{3}}}}=-a_{1}+\frac{1}{2}a_{2}=3.88,
\label{eq:GH}
\end{align}
where numerical values have been calculated with the help of Eq.~(\ref{eq:a123}).
With these definitions of the couplings the formulas for the decay
widths averaged over the initial isospin and summed over the final
isospin read as follows:% 
\begin{align}
\Gamma_{\Sigma(\boldsymbol{6}_{1})\rightarrow\Lambda(\overline{\boldsymbol{3}%
}_{0})+\pi}  &  =\frac{1}{72\pi}\frac{p^{3}}{F_{\pi}^{2}}\frac{M_{\Lambda
(\overline{\boldsymbol{3}}_{0})}}{M_{\Sigma(\boldsymbol{6}_{1})}}%
H_{{\overline{\boldsymbol{3}}}}^{2}\frac{3}{8},\nonumber\\
\Gamma_{\Xi(\boldsymbol{6}_{1})\rightarrow\Xi(\overline{\boldsymbol{3}}%
_{0})+\pi}  &  =\frac{1}{72\pi}\frac{p^{3}}{F_{\pi}^{2}}\frac{M_{\Xi
(\overline{\boldsymbol{3}}_{0})}}{M_{\Xi(\boldsymbol{6}_{1})}}H_{{\overline
{\boldsymbol{3}}}}^{2}\frac{9}{32},\nonumber\\
\Gamma_{\Omega(\overline{\boldsymbol{15}}_{1})\rightarrow\Xi(\overline
{\boldsymbol{3}}_{0})+K}  &  =\frac{1}{72\pi}\frac{p^{3}}{F_{K}^{2}}%
\frac{M_{\Xi(\overline{\boldsymbol{3}}_{0})}}{M_{\Omega(\overline
{\boldsymbol{15}}_{1})}}G_{{\overline{\boldsymbol{3}}}}^{2}\frac{3}%
{10},\nonumber\\
\Gamma_{\Omega(\overline{\boldsymbol{15}}_{1})\rightarrow\Omega(\boldsymbol{6}%
_{1})+\pi}  &  =\frac{1}{72\pi}\frac{p^{3}}{F_{\pi}^{2}}\frac{M_{\Omega
(\boldsymbol{6}_{1})}}{M_{\Omega(\overline{\boldsymbol{15}}_{1})}%
}G_{\boldsymbol{6}}^{2}\frac{4}{15},\nonumber\\
\Gamma_{\Omega(\overline{\boldsymbol{15}}_{1})\rightarrow\Xi(\boldsymbol{6}%
_{1})+K}  &  =\frac{1}{72\pi}\frac{p^{3}}{F_{K}^{2}}\frac{M_{\Xi
(\boldsymbol{6}_{1})}}{M_{\Omega(\overline{\boldsymbol{15}}_{1})}%
}G_{\boldsymbol{6}}^{2}\frac{2}{15}. \label{eq:widths}%
\end{align}
When we need a decay width for a specific isospin combination, the
widths of Eqs.~(\ref{eq:widths}) have to be multiplied by a pertinent SU(2)
Clebsch-Gordan coefficient.

Note that in the $\chi$QSM couplings (\ref{eq:a123}) are expressed in terms of
the inertia parameters \cite{Praszalowicz:1998jm},
\begin{equation}
a_{1}=A_{0}-\frac{B_{1}}{I_{1}},\;a_{2}=2\frac{A_{2}}{I_{2}},\;a_{3}%
=2\frac{A_{1}}{I_{1}}.
\end{equation}
Since all inertia parameters scale as $N_{c}$, we see that formally $a_{1}$
contains both leading and the first subleading term in $N_{c}$, whereas
$a_{2,3}$ scale as $N_{c}^{0}$. This can be the best seen in the
nonrelativistic (NR) limit for small soliton size, where
\cite{Praszalowicz:1995vi}
\begin{equation}
A_{0}\rightarrow-N_{c},\;\frac{B_{1}}{I_{1}}\rightarrow2,\;,\;\frac{A_{2}%
}{I_{2}}\rightarrow2,\;\frac{A_{1}}{I_{1}}\rightarrow1 \label{nrel}%
\end{equation}
or%
\begin{equation}
a_{1}\rightarrow-(N_{c}+2),\;a_{2}\rightarrow4,\;a_{3}\rightarrow2
\label{eq:NRa123}%
\end{equation}
Remember that in this limit the axial-vector coupling constant% 
\begin{equation}
g_{A}=\frac{7}{30}\left(  -a_{1}+\frac{1}{2}a_{2}+\frac{1}{14}a_{3}\right)
\rightarrow\frac{5}{3},
\end{equation}
which is equal to the naive quark model result for $g_{A}$, whereas for the
phenomenological values (\ref{eq:a123}) $g_{A}=1.23.$

In the case of heavy baryons all inertia parameters should be
rescaled by approximately\footnote{Strictly speaking rescaling by a
  factor $(N_c -1)/N_c$ should work well only for quantities dominated
  by valence levels. As the contribution of the sea quarks to some
  quantities can be sizeable one may expect 10\%--20\% variations of that
  rescaling factor.} $(N_{c}-1)/N_c$, 
which does not change the scaling of their ratios; however it does
change the value of $a_{1}$(strictly speaking the $A_{0}$ part of
$a_{1}$). Therefore for heavy baryons we should use%
\begin{equation}
A_{0}\rightarrow\tilde{A}_{0}=\frac{N_c-1}{N_c}A_{0}. \label{ea:A0scaled}%
\end{equation}
Unfortunately from the fits to the experimental data we do not know separately
the values of $A_{0}$ and $B_{1}/I_{1}$. We know only these values in the NR
limit (\ref{nrel}). Making the rather bold assumption that NR 
relation $B_{1}=2A_{1}$ holds also for realistic soliton sizes, we 
approximate $B_{1}/I_{1}\sim a_{3}$, which gives 
$A_{0}=a_{1}+a_{3}$. Therefore, following Eq.(\ref{ea:A0scaled}), we
make the following replacement:
\begin{equation}
a_{1}\rightarrow\tilde{a}_{1}=\left[  \frac{N_c-1}{N_c}\left(  a_{1}+a_{3}\right)
-a_{3}\right]  \sigma. \label{eq:a1scaled}%
\end{equation}
Here $\sigma$ is a correction factor that takes into account possible
deviations from the assumption that $B_{1}=2A_{1}$ and possible deviation of the 
rescaling factor from $(N_c-1)/N_c$. 
The parameter $\sigma$ characterizes the modification of the mean
field when one goes from $N_c$ quarks in light baryons to $N_c-1$
quarks in heavy baryons. In the ideal case $\sigma$ should be close to
unity; however in practice, as we shall see in the following, a $15\%$
correction is required to get satisfactory description of the decay
widths. Such small modification of the mean field is fully compatible
with the expected size of $1/N_c$ corrections. 

It is interesting to calculate the decay constants (\ref{eq:GH}) in the nonrelativistic
quark model limit (\ref{eq:NRa123}) with $a_{1}\rightarrow-(N_{c}+1)$. One can see
that in this limit $G_{\boldsymbol{6}}=0$. This means that the decay channels of
the putative heavy pentaquarks to the sextets should be strongly suppressed. This situation is
very similar to the suppression of $\Theta ^{+}(1530)$ decay
width~\cite{Diakonov:1997mm}. 

One should remember that in the large $N_c$ limit flavor representations of  the light sector
should be generalized \cite{Karl:1985qy,Dulinski:1987er,Piesciuk:2007xq}, and in the present case the standard 
generalization takes the following form \cite{QlargeNc}:
\begin{align}
"\boldsymbol{\overline{3}}"  &  =(0,q+1),\nonumber\\
"\boldsymbol{6}"  &  =(2,q), \nonumber\\
"\overline{\boldsymbol{15}}"  &  =(1,q+2)
\end{align}
with $q=(N_c-3)/2$. With this generalization the pertinent SU(3)
Clebsch-Gordan coefficients acquire $N_c$
dependence. For example
\begin{equation}
G_{\overline{\boldsymbol{3}}}=-a_{1}-\frac{N_c-1}{4}a_{2},
\end{equation}
which means that $G_{\overline{\boldsymbol{3}}}$ is suppressed in
the large $N_c$  NR limit  (\ref{eq:NRa123}), since the leading $N_c$ terms cancel out.
This cancellation is similar as in the case of high-dimensional exotic representations of
light baryons \cite{Piesciuk:2007xq}. 
We see that large $N_c$ counting together with the arguments based on the nonrelativistic
limit explain the numerical hierarchy of the decay couplings (\ref{eq:GH}).
Full $N_c$ dependence of the mass splittings
and decay widths will be discussed elsewhere   \cite{QlargeNc}.

\section{Decay widths of sextet heavy baryons}
\label{sec:sextetdecays}

A number of decay widths are measured for heavy baryons. In
Ref.~\cite{Cheng:2006dk} decays of charm sextet $\Sigma_{c}$, both of
spin 1/2 and 3/2, have been fitted with the help of the formula
analogous to (\ref{eq:widths}) in terms of a single coupling
$g_{2}$. The updated phenomenological value is presently $g_{2}=0.56$
with a 5\% error. Coupling $g_{2}$ can be expressed in terms of
$H_{{\overline{\boldsymbol{3}}}}$:% 
\begin{equation}
\left\vert g_{2}\right\vert
=\frac{1}{4\sqrt{3}}H_{{\overline{\boldsymbol{3}}% 
}}=\frac{1}{4\sqrt{3}}\left(  -\tilde{a}_{1}+\frac{1}{2}a_{2}\right)  .
\end{equation}
To fit the phenomenological value of 0.56 we need a correction factor
$\sigma=0.85$. Other than that there is no extra freedom and all decay widths
are genuine predictions of the present model.

Two comments are in order here. Decay widths of different $T_{3}$
states show small isospin violations, which are mainly due to a small
phase volume and hence due to the sensitivity to the mass difference of
$\pi^{\pm }-\pi^{0}$. Decay couplings are calculated in the isospin
symmetric limit. Secondly, different decays of particles within the
same initial and final SU(3) multiplets
($\mathcal{R}_{1}\rightarrow\mathcal{R}_{2}$) are related by the SU(3)
symmetry, which is more general than the present model. $\chi$QSM
relates couplings of different SU(3) multiplets, like $H_{{\overline
    {\boldsymbol{3}}}}$, $G_{\overline{\boldsymbol{3}}}$ or
$G_{\boldsymbol{6}}$, which are expressed as different combinations of
$a_{1,2,3}$ (\ref{eq:a123}). 

Below, in Tables \ref{tab:charm} and \ref{tab:bottom}, we list our
results and the experimental values both for the charm and bottom
baryons. Note that in the case of
$\Xi_{Q}^{q}(\boldsymbol{6}_{1},3/2)$ (where $q$ denotes the 
pertinent charge) decays to $\Xi_{Q}(\overline{\boldsymbol{3}}_{0},1/2)+\pi$
the width is summed with the CG weighting $\frac{1}{3}(\Xi_{Q}^{q}%
(\overline{\boldsymbol{3}}_{0},1/2)+\pi^{0})+\frac{2}{3}(\Xi_{Q}^{q\mp
1}(\overline{\boldsymbol{3}}_{0},1/2)+\pi^{\pm})$. When no charges are
specified the width is averaged over initial and summed over final isospin.

\begin{table}[h]
\centering{}
\begin{tabular}
[c]{|r|c|r|c|}\hline
\# & \text{Decay} & $%
\begin{array}
[c]{c}%
\text{This}\\
\text{work}%
\end{array}
$ & \text{Exp.}\\\hline
$1$ & $\Sigma_{c}^{++}(\boldsymbol{6}_{1},1/2)\rightarrow\Lambda_{c}%
^{+}(\overline{\boldsymbol{3}}_{0},1/2)+\pi^{+}$ & $1.93 $ & $1.89_{-0.18}%
^{+0.09}$\\
$2 $ & $\Sigma_{c}^{+}(\boldsymbol{6}_{1},1/2)\rightarrow\Lambda_{c}%
^{+}(\overline{\boldsymbol{3}}_{0},1/2)+\pi^{0} $ & $2.24 $ & $<4.6$\\
$3 $ & $\Sigma_{c}^{0}(\boldsymbol{6}_{1},1/2)\rightarrow\Lambda_{c}%
^{+}(\overline{\boldsymbol{3}}_{0},1/2)+\pi^{-} $ & $1.90 $ & $1.83_{-0.19}%
^{+0.11}$\\\hline
$4 $ & $\Sigma_{c}^{++}(\boldsymbol{6}_{1},3/2)\rightarrow\Lambda_{c}%
^{+}(\overline{\boldsymbol{3}}_{0},1/2)+\pi^{+} $ & $14.47 $ & $14.78_{-0.19}%
^{+0.30}$\\
$5 $ & $\Sigma_{c}^{+}(\boldsymbol{6}_{1},3/2)\rightarrow\Lambda_{c}%
^{+}(\overline{\boldsymbol{3}}_{0},1/2)+\pi^{0} $ & $15.02  $ & $<17$\\
$6 $ & $\Sigma_{c}^{0}(\boldsymbol{6}_{1},3/2)\rightarrow\Lambda_{c}%
^{+}(\overline{\boldsymbol{3}}_{0},1/2)+\pi^{-} $ & $14.49  $ & $15.3_{-0.5}%
^{+0.4}$\\\hline
$7 $ & $\Xi_{c}^{+}(\boldsymbol{6}_{1},3/2)\rightarrow\Xi_{c}(\overline
{\boldsymbol{3}}_{0},1/2)+\pi$ & $2.35  $ & $2.14\pm0.19$\\
$8 $ & $\Xi_{c}^{0}(\boldsymbol{6}_{1},3/2)\rightarrow\Xi_{c}(\overline
{\boldsymbol{3}}_{0},1/2)+\pi$ & $2.53 $ & $2.35\pm0.22$\\\hline
\end{tabular}
\caption{Charm sextet baryons decay widths in MeV. Experimental data are taken from Particle Data Group \cite{Patrignani:2016xqp}.}%
\label{tab:charm}%
\end{table}

\begin{table}[h]
\centering{}
\begin{tabular}
[c]{|r|c|r|c|}\hline
\# & \text{Decay} & $%
\begin{array}
[c]{c}%
\text{This}\\
\text{work}%
\end{array}
$ & \text{Exp.}\\\hline
$1$ & $\Sigma_{b}^{+}(\boldsymbol{6}_{1},1/2)\rightarrow\Lambda_{b}%
^{0}(\overline{\boldsymbol{3}}_{0},1/2)+\pi^{+} $ & $6.12 $ & $9.7_{-3.0}%
^{+4.0}$\\
$2$ & $\Sigma_{b}^{-}(\boldsymbol{6}_{1},1/2)\rightarrow\Lambda_{b}%
^{0}(\overline{\boldsymbol{3}}_{0},1/2)+\pi^{-} $ & $6.12 $ & $4.9_{-2.4}%
^{+3.3}$\\\hline
$3$ & $\Xi_{b}^{^{\prime}}(\boldsymbol{6}_{1},1/2)\rightarrow\Xi_{c}%
(\overline{\boldsymbol{3}}_{0},1/2)+\pi$ & $0.07 $ & $<0.08$\\\hline
$4 $ & $\Sigma_{b}^{+}(\boldsymbol{6}_{1},3/2)\rightarrow\Lambda_{b}%
^{0}(\overline{\boldsymbol{3}}_{0},1/2)+\pi^{+} $ & $10.96 $ & $11.5\pm2.8$\\
$5 $ & $\Sigma_{b}^{-}(\boldsymbol{6}_{1},3/2)\rightarrow\Lambda_{c}%
^{0}(\overline{\boldsymbol{3}}_{0},1/2)+\pi^{-} $ & $11.77 $ & $7.5\pm
2.3$\\\hline
$6 $ & $\Xi_{b}^{0}(\boldsymbol{6}_{1},3/2)\rightarrow\Xi_{b}(\overline
{\boldsymbol{3}}_{0},1/2)+\pi$ & $0.80 $ & $0.90\pm0.18$\\
$7 $ & $\Xi_{b}^{-}(\boldsymbol{6}_{1},3/2)\rightarrow\Xi_{b}(\overline
{\boldsymbol{3}}_{0},1/2)+\pi$ & $1.28 $ & $1.65\pm0.33$\\\hline
\end{tabular}
\caption{Bottom sextet baryons decay widths in MeV. Experimental data are taken from Particle Data Group \cite{Patrignani:2016xqp}.}%
\label{tab:bottom}%
\end{table}

We see remarkably good agreement of the $\chi$QSM results with the
experimental widths for {\it both} charm and bottom baryons. 
To better illustrate this we have plotted in Figs.~\ref{fig:charmwidths}
and \ref{fig:bottomwidths} the results collected in
Tables \ref{tab:charm} and \ref{tab:bottom}.
We stress
that for the calculation of the heavy baryon widths essentially we did
not need any new parameter -- everything was fixed in the light baryon
sector. The scaling factor $\sigma$ introduced in
Eq.~(\ref{eq:a1scaled}) is not really a new parameter; it rather
characterizes the $\sim$15~\% modification of the universal meson mean
field due to $1/N_c$ corrections.   

\section{Decay widths of exotic antidecapentaplet baryons}
\label{sec:a15decays}

With all decay constants fixed (\ref{eq:a123}) from the decays of {\it
  light} baryons we can now predict the widths of the putative
pentaquark $\Omega_{c}(3050)$ and $\Omega_{c}(3119)$ states that
are collected in Tables \ref{tab:Omc_a} and \ref{tab:Omc_b}. Note that
the exotic $\Omega_c$'s from $\overline{\boldsymbol{15}}$ 
have the isospin one and hence the decay mode to $\Omega_c(\boldsymbol{6})+\pi$ is
allowed. However, the corresponding decay constant 
$G_6=-\tilde{a}_{1}-\frac{1}{2}a_{2}-a_{3}$ is zero in the
nonrelativistic small soliton size limit, analogously as for the
corresponding coupling of the light pentaquark $\Theta^+$. Therefore,
we expect that the $\Omega_c(\boldsymbol{6})+\pi$ decay mode should be
strongly suppressed. 

\begin{table}[h]
\centering{}
\begin{tabular}
[c]{|r|c|c|c|}\hline
\# & \text{Decay} & $%
\begin{array}
[c]{c}%
\text{This}\\
\text{work}%
\end{array}
$ & Exp.\\\hline
& $\Omega_{c}(\overline{\boldsymbol{15}}_{1},1/2)\rightarrow\Xi_{c}%
(\overline{\boldsymbol{3}}_{0},1/2)+K $ & 0.339 & $-$\\
& $\Omega_{c}(\overline{\boldsymbol{15}}_{1},1/2)\rightarrow\Omega
_{c}(\boldsymbol{6}_{1},1/2)+\pi$ & 0.097 & $-$\\
& $\Omega_{c}(\overline{\boldsymbol{15}}_{1},1/2)\rightarrow\Omega
_{c}(\boldsymbol{6}_{1},3/2)+\pi$ & 0.045  & $-$\\\hline
9 & \text{Total} & 0.48  & $0.8\pm0.2\pm0.1$\\\hline
\end{tabular}
\caption{$\Omega_{c}(\overline{\boldsymbol{15}}_{1},1/2)$ partial and total
decay widths in MeV. Experimental value is from the LHCb measurement \cite{Aaij:2017nav}.}%
\label{tab:Omc_a}%
\end{table}

\begin{table}[tb]
\centering{}
\begin{tabular}
[c]{|r|c|c|c|}\hline
\# & \text{Decay} & $%
\begin{array}
[c]{c}%
\text{This}\\
\text{work}%
\end{array}
$ & Exp.\\\hline
& $\Omega_{c}(\overline{\boldsymbol{15}}_{1},3/2)\rightarrow\Xi_{c}%
(\overline{\boldsymbol{3}}_{0},1/2)+K$ & 0.848 & $-$\\
& $\Omega_{c}(\overline{\boldsymbol{15}}_{1},3/2)\rightarrow\Xi_{c}%
(\boldsymbol{6}_{1},1/2)+K$ & 0.009 & $-$\\
& $\Omega_{c}(\overline{\boldsymbol{15}}_{1},3/2)\rightarrow\Omega
_{c}(\boldsymbol{6}_{1},1/2)+\pi$ & 0.169 & $-$\\
& $\Omega_{c}(\overline{\boldsymbol{15}}_{1},3/2)\rightarrow\Omega
_{c}(\boldsymbol{6}_{1},3/2)+\pi$ & 0.096 & $-$\\\hline
10 & \text{Total} & 1.12 & $1.1\pm0.8\pm0.4$\\\hline
\end{tabular}
\caption{$\Omega_{c}(\overline{\boldsymbol{15}}_{1},3/2)$ partial and total
decay widths in MeV. Experimental value is from the LHCb measurement \cite{Aaij:2017nav}}%
\label{tab:Omc_b}%
\end{table}

Note that the kaon momentum in the decay of $\Omega_c(3050)$,
  $p_K=275$~MeV, is quite close to the pion momentum in the decay of
  $\Delta$, $p_{\pi}=228$~MeV, yet the $\Delta$ decay width is two
  orders of magnitude larger than the one of $\Omega_c(3050)$. 
From  Tables~\ref{tab:Omc_a} 
and~\ref{tab:Omc_b}  we see that, despite the large
phase volume available, the predicted decay widths are very small and are
in agreement with the small ($\sim 1$~MeV) decay widths observed by
the LHCb collaboration (see also Fig.~\ref{fig:charmwidths}). Note
that $\Omega_c(3050)$ and $\Omega_c(3119)$ are the narrowest states
among five LHCb $\Omega_c$'s and our mean field picture gives a natural,
parameter-free explanation of this observation. 
In the large $N_c$ nonrelativistic 
limit discussed at the end of Sec.~\ref{sec:genformalism} the decay constant
to $\boldsymbol{6}$ is strongly suppressed,
whereas the decay constant to $\overline{\boldsymbol{3}}$ is suppressed
in the leading order of $N_c$.

\begin{figure}[h]
\centering
\includegraphics[height=7cm]{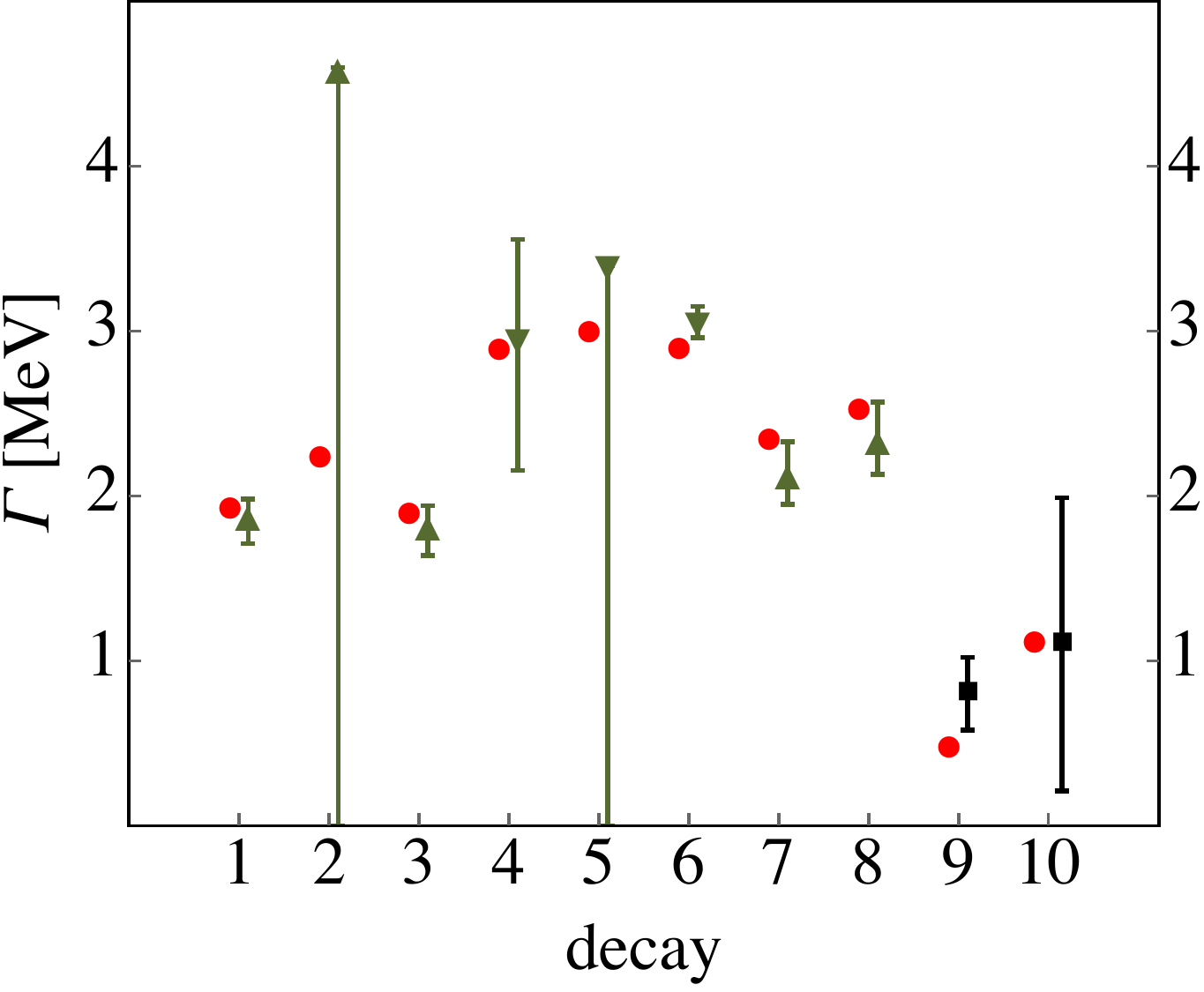} 
\caption{Decay widths of the
charm baryons. Numbers on the horizontal axis label the decay modes as
listed in Tables \ref{tab:charm}, \ref{tab:Omc_a} and
\ref{tab:Omc_b}. Red full circles correspond to our theoretical
predictions. Dark green triangles correspond to the experimental data
\cite{Patrignani:2016xqp}. Data for decays 4 -- 7 of 
$\Sigma_{c}(\boldsymbol{6}_{1},3/2)$ (down-triangles) have been
divided by a factor of 5 to fit within the plot area. Widths of two
LHCb \cite{Aaij:2017nav} $\Omega_{c}$ states that we interpret as pentaquarks are plotted
as black full squares. }% 
\label{fig:charmwidths}%
\end{figure}

\begin{figure}[h]
\centering
\includegraphics[height=7cm]{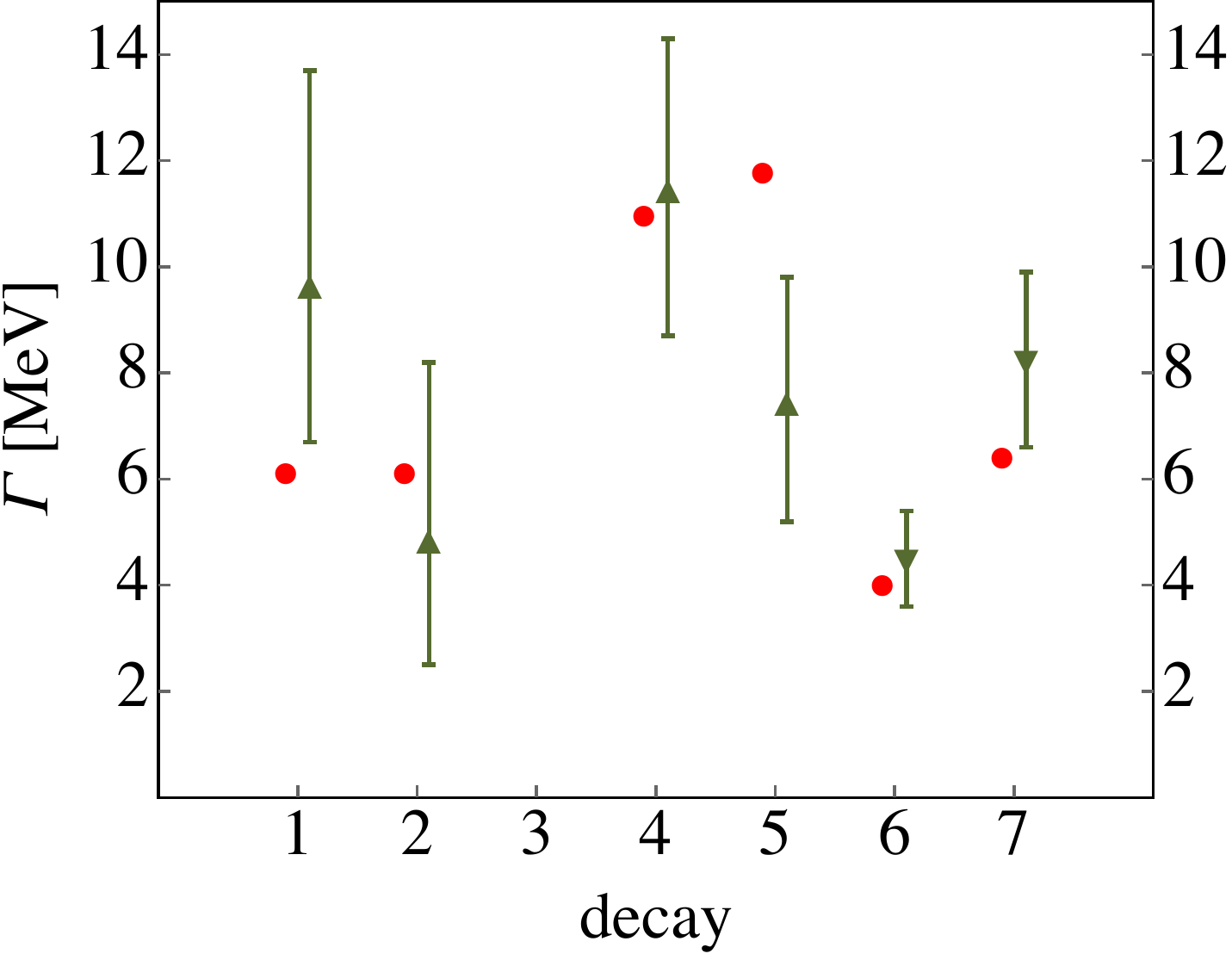} 
\caption{Decay widths of
the bottom baryons. Numbers on the horizontal axis label the decay
modes as listed in Table \ref{tab:bottom}. Red full circles correspond
to our theoretical predictions. Dark green triangles correspond to the
experimental data \cite{Patrignani:2016xqp}. Data for decays 6 and 7
of $\Xi_{b}(\boldsymbol{6}_{1},3/2)$ (down-triangles) have been
multiplied by a factor of 5 to be better visible on the plot. }% 
\label{fig:bottomwidths}%
\end{figure}

The results in this section also imply that other members of the exotic
antidecapentaplet are expected to be anomalously narrow. All their 
partial decay widths can be easily computed in our model
with the help of general
formula (\ref{eq:Gammageneral}). As an illustration
we quote here the result for the decay widths of other explicitly
exotic members of the antidecapentaplet, $\Xi_c^{3/2-}$ and
$\Xi_c^{3/2++}$, which have the minimal quark content ($cdds\bar u$)
and ($cuus\bar d$). The masses of these states are predicted in
Ref.~\cite{Kim:2017jpx} to be 2931 and 3000~MeV for the
$J^P= 1/2^+$ and $J^P= 3/2^+$ multiplets, respectively. 
The predictions for the partial widths\footnote{Note that the decay
  $\Xi_c^{3/2}\to \Lambda_c+K$ is forbidden by the isospin symmetry.}
of exotic $\Xi_c^{3/2}$ are given in Table~\ref{tab:Xic_a}. We see
that, indeed, the widths are anomalously small. Interesting is that
the width of the isospin-1/2 $\Xi_c$ from
$\overline{\boldsymbol{15}}$-plet is even smaller ($ <1$~MeV), 
with the dominant decay mode: $\Lambda_c+K$. 

\begin{table}[h]
\centering{}
\begin{tabular}
[c]{|c|c|c|}\hline
 \text{Decay} & 
$J= 1/2$ &
  $J= 3/2$\\\hline
 $\Xi^{3/2}_{c}(\overline{\boldsymbol{15}}_{1},J)\rightarrow\Xi_{c}%
(\overline{\boldsymbol{3}}_{0},1/2)+\pi $ & 1.67 & 2.49 \\
 $\Xi^{3/2}_{c}(\overline{\boldsymbol{15}}_{1},J)\rightarrow\Xi
_{c}(\boldsymbol{6}_{1},1/2)+\pi$ & 0.045 & 0.079 \\
 $\Xi^{3/2}_{c}(\overline{\boldsymbol{15}}_{1},J)\rightarrow\Xi
_{c}(\boldsymbol{6}_{1},3/2)+\pi$ & 0.022 & 0.046\\
 $\Xi^{3/2}_{c}(\overline{\boldsymbol{15}}_{1},J)\rightarrow\Sigma
_{c}(\boldsymbol{6}_{1},1/2)+K$ & $-$ & 0.019\\\hline
   \text{Total} & 1.74 & 2.64 \\\hline
\end{tabular}
\caption{Predictions in MeV for the partial and total
decay widths of explicitly exotic
$\Xi^{3/2}_{c}(\overline{\boldsymbol{15}}_{1},J)$.}% 
\label{tab:Xic_a}%
\end{table}

\section{Conclusions}
\label{sec:con}

In the large $N_c$ limit both heavy and light baryons are described by
the universal mean field. This allows us to relate the properties of
heavy baryons to the light ones. The goal of the present paper was twofold:
first, to test the mean-filed picture of baryons against the data on
strong decays of sextets of charm and bottom baryons. Second, to make
predictions for the decay widths of exotic antidecapentaplet
($\overline{\boldsymbol{15}}$) baryons.  

With the only input from the decays of light octet baryons (due to the
universality of the chiral mean field) we have obtained 
excellent description of strong decays of {\it both} charm and bottom
sextets.  The agreement is illustrated in Fig.~\ref{fig:charmwidths}
for charm baryon decays and in Fig.~\ref{fig:bottomwidths} for bottom 
decays. We have also shown that going from $N_c$ light quarks in light
baryons to  $N_c-1$ light quarks in heavy baryons the mean 
field is modified by about 15\%. That moderate modification is an
agreement with expected size of $1/N_c$ corrections. 

Given the excellent agreement of our calculation with the measured widths
of ground-state heavy baryons we made parameter-free predictions for
the decays of exotic $\overline{\boldsymbol{15}}$-plet baryons. We
have shown that the widths of $\overline{\boldsymbol{15}}$ baryons
must be anomalously small, due to essentially the same mechanism as
in the case of narrow anti-decuplet  light baryons. In
particular, for $\Omega_c(3050)$ and $\Omega_c(3119)$, which we
interpreted in Ref.~\cite{Kim:2017jpx} as belonging to the
antidecapentaplet,  we obtained widths of  0.48~MeV and 1.12~MeV
correspondingly. Experimentally, these two (among five) states have
the smallest widths \cite{Aaij:2017nav} which are in agreement with
our parameter-free calculations, see Tables~\ref{tab:Omc_a} and
\ref{tab:Omc_b} and Fig.~\ref{fig:charmwidths}. 
The parametrical
suppression of the pertinent  decay constants has been discussed
at the end of Sec.~\ref{sec:genformalism}. We have shown that 
theoretical arguments based on large $N_c$ and NR limits explain
the numerical hierarchy of the  decay couplings (\ref{eq:GH}).

For the complete description of the LHCb $\Omega_c$ states, estimates
of strong decays for negative parity baryons are needed. In the
$\chi$QSM negative parity baryons correspond 
to the configuration with one quark excited from the Dirac see to the empty 
valence level (see Ref.~\cite{Kim:2017jpx}). Therefore one expects that
the decay operator will depend on a new parameter related to this transition,
similarly to the case of mass splittings for these states. Furthermore, 
negative parity particles will have $s$-wave and/or $d$-wave decays,
which are not possible if the parity is positive. Finally, for heavier states,
decays to light baryons and heavy mesons are possible. Such decays
require new theoretical treatment within the framework of the $\chi$QSM
since the Goldberger-Treiman relation is not directly applicable in this case.
All these issues require further study therefore we have not
attempted to address them in the present paper. 

The results of the present study reinforce our conclusions from
Ref.~\cite{Kim:2017jpx} that the two narrowest $\Omega^0_c$ states
reported recently by the LHCb collaboration in Ref.~\cite{Aaij:2017nav}
correspond to the exotic SU(3) multiplet, namely the antidecapentaplet
($\overline{\boldsymbol{15}}$).  As seen from Fig.~\ref{fig:irreps2}
these states belong to the isospin triplet, rather than the
singlet. Therefore this quantum number assignment can be relatively
easily verified experimentally.  

\acknowledgements
H.-Ch.~K. is grateful to J.Y. Kim for discussion. The present work
was supported by Basic Science Research Program through
the National Research Foundation of Korea funded by the Ministry of
Education, Science and Technology (Grant No.
NRF-2015R1A2A2A04007048 (H.-Ch.~K.) and Grant No.
NRF-2016R1C1B1012429 (Gh.-S.~Y.)). The work of M.V.P. is supported by  
Sino-German Collaborative Research Centre~110. M.V.P. is also grateful to  the
Department of Theoretical Physics of Irkutsk State University for
hospitality and support. The work of M.P. has been performed in the framework
of COST Action CA15213 THOR.

%\pagebreak

\end{document}